\documentclass[11pt,draftcls,onecolumn]{IEEEtran}

\usepackage{multicol}
\usepackage{indentfirst,flushend}
\usepackage{amssymb,bm,mathrsfs,bbm,amscd,amsmath,amsthm}
\usepackage{graphicx}
\usepackage{color}
\usepackage{geometry,cases}

\newtheorem{lemma}{Lemma}
\newtheorem{remark}{Remark}
\usepackage{multirow}
\usepackage{cite}
\usepackage{citesort}
\usepackage{stfloats}
\usepackage[nooneline,flushleft]{caption2}
\usepackage{algorithm,clrscode}
\usepackage{algorithmic}

\begin{document}

\title{Max-Min Energy Efficient Beamforming for Multicell Multiuser Joint Transmission Systems}

\author{Shiwen~He, Yongming~Huang,~\IEEEmembership{Member,~IEEE}, Shi~Jin,~\IEEEmembership{Member,~IEEE}, Fei~Yu, and Luxi~Yang,~\IEEEmembership{Member,~IEEE}
\thanks{S. He, Y. Huang, S. Jin, F. Yu, and L. Yang are with the School of Information Science and Engineering, Southeast University, Nanjing 210096, China. (Email:\{hesw01, huangym, jinshi, yufei, lxyang\}@seu.edu.cn).

}
}

\maketitle

\maketitle \vspace{-.6 in}

\begin{abstract}
Energy efficient communication technology has attracted much attention due to the explosive growth of energy consumption in current wireless communication systems. In this letter we focus on fairness-based energy efficiency and aim to maximize the minimum user energy efficiency in the multicell multiuser joint beamforming system, taking both dynamic and static power consumptions into account. This optimization problem is a non-convex fractional programming problem and hard to tackle. In order to find its solution, the original problem is transformed into a parameterized polynomial subtractive form by exploiting the relationship between the user rate and the minimum mean square error, and using the fractional programming theorem. Furthermore, an iterative algorithm with proved convergence is developed to achieve a near-optimal performance. Numerical results validate the effectiveness of the proposed solution and show that our algorithm significantly outperforms the max-min rate optimization algorithm in terms of maximizing the minimum energy efficiency.
\end{abstract}


\section*{\sc \uppercase\expandafter{\romannumeral1}. Introduction}

Multiple-input multiple-output (MIMO) technology is maturing and is being incorporated into emerging advanced wireless communication systems owing to their potential of significantly improving the spectral efficiency~\cite{MagGesbert2007,TWCDahrouj2010,TSPBhagavatula2011,MagRusek2013}. In the past few years, multiple-point coordinated or joint transmission, or network MIMO, has also attracted extensive concerns and has been widely studied. However, it should be noted that most of existing researches focused on maximizing the capacity of the wireless communication or balancing the user rates subject to given power constraints~\cite{JSACGESbert2010,JSACZhang2010,TSPBogale2012,TWCHuang2012}. In particular, a cooperative multicell block diagonalization joint transmission scheme was proposed in~\cite{JSACZhang2010}, considering per-base station (BS) power constraints. Centralized and distributed algorithms which aim to maximize the weighted sum rate were developed in~\cite{TSPBogale2012}. Recently, Huang \emph{etc al} proposed a distributed algorithm to maximize the minimum signal-to-interference-noise ratio (SINR) for coordinated beamforming systems~\cite{TWCHuang2012}.

More recently, green radio or energy efficient communication has drawn increasing attention~\cite{SurverFeng2013,TWCLe2013,LetterMao2013,TWCNg201209}. In~\cite{TWCLe2013} an energy efficient multiuser MIMO transmission was designed to maximize the system energy efficiency (EE) which was defined as the ratio of the sum rate to the total power consumption. Energy efficient optimization for cognitive radio MIMO broadcast channels was also studied subject to the total power constraint, the interference power constraint and the minimum system throughput constraint~\cite{LetterMao2013}. In \cite{TWCLe2013} and \cite{LetterMao2013}, the EE optimization problem was solved by applying the multiple access channel broadcast duality theory and employing dirty paper coding (DPC). Besides, energy efficient resource allocation has been  studied for orthogonal frequency division multiple access (OFDMA) downlink systems with a large number of transmit antennas and fixed beamformer~\cite{TWCNg201209}.

Contrary to these existing literature which focused on maximizing the system EE~\cite{SurverFeng2013,TWCLe2013,LetterMao2013,TWCNg201209}, in this letter we study a fairness-based EE problem, i.e., to maximize the minimum user EE which is defined as the ratio of the user rate to its power consumption. Note that this new criterion could guarantee the EE of each individual node, which is particulary important for heterogenous networks where some nodes may have stringent EE requirement. However, this optimization problem of interest is nonconvex fractional problem and therefore difficult to solve directly. To address it, the user rate is firstly reexpressed as an equivalent optimization form with additional auxiliary variables by exploiting the relationship between the user rate and the minimum mean square error (MMSE)~\cite{ICCJose2011}. Then, the originally fractional problem is transformed into a parameterized quadratic subtractive form using the fractional theorem~\cite{JstorJagan1966,MathCrouzeix1991}. Based on that, an iterative algorithm with guaranteed convergence is proposed to solve the fairness-based EE problem.

\section*{\sc \uppercase\expandafter{\romannumeral2}. System Model}

We consider a multicell multiuser joint transmission system consisting of $K$ cells, each of which has one BS equipped with $\widetilde{M}$ antennas and $\widetilde{N}$ single-antenna users. Allowing full cooperation between the BSs, i.e., they perform joint transmission to users, the cooperative multicell downlink system can be modeled as a super MISO broadcast channel (BC) with $M$ transmit antennas and $N$ users, where $M=K\widetilde{M}$, $N=K\widetilde{N}$. For convenience, we assign the antenna indices according to the BS index, i.e., the $\left(\left(k-1\right)\widetilde{M}+1\right)$-th to $\left(k\widetilde{M}+1\right)$-th antennas represent the $\widetilde{M}$ antennas from the $k$-th BS, $\forall k$. Similarly, the indices of MSs in the super MISO BC are assigned according to their cell indices, i.e., the $\left(\left(k-1\right)\widetilde{N}+1\right)$-th to $\left(k\widetilde{N}+1\right)$-th users represent the $\widetilde{N}$ users from $k$-th cell. Then, the received signal of the $n$-th user is denoted as
\begin{equation}\label{Fairness_Energy_Efficiency_1}
y_{n}=\sum_{m=1}^{N}\bm{h}_{n}^{H}\bm{w}_{m}x_{m}+z_{n}
\end{equation}
where $\bm{h}_{n}=[\bm{h}_{n,1}^{T},\cdots,\bm{h}_{n,K}^{T}]^{T}\in\mathbb{C}^{M}$ denotes the flat channel fading coefficient from all the $M$ BS antennas to the $n$-th user, including both the large scale fading and the small scale fading,  $\bm{w}_{n}$ denotes the beamforming vector for the $n$-th user, $x_{n}$ denotes the transmitted signal for the $n$-th user with zero mean and unit variance, and $z_{n}$ denotes the additive white Gaussian noise with zero mean and variance $\sigma_{n}^{2}$.

\section*{\sc \uppercase\expandafter{\romannumeral3}. Optimization Target Formulations}

Different from the conventional EE criterion which is defined as the ratio of the system sum rate to the total power consumption~\cite{SurverFeng2013,TWCLe2013,LetterMao2013,TWCNg201209}, the criterion of interest is individual user EE defined as the ratio of the user rate to the user power consumption, given by
\begin{equation}\label{Fairness_Energy_Efficiency_2}
f_{n}\left(\{\bm{w}_{n}\}\right)=\frac{r_{n}}{\left\|\bm{w}_{n}\right\|^{2}+\frac{MP_{c}+KP_{0}}{N}}
\end{equation}
where $P_{c}$ is the constant circuit power consumption per antenna which are independent of the actual transmitted power, $P_{0}$ is the basic power consumed at the BS independent of the number of transmit antennas, and $r_{n}$ denotes the instantaneous rate of the $n$-th user and is calculated as $r_{n}=\log \left(1+\mbox{SINR}_{n}\right)$, in unit of Nat/s/Hz, where $\mbox{SINR}_{n}$ denotes the SINR of the $n$-th user and is expressed as
\begin{equation}\label{Fairness_Energy_Efficiency_4}
\mbox{SINR}_{n}=\frac{\left|\bm{h}_{n}^{H}\bm{w}_{n}\right|^{2}}
{\sum\limits_{m=1, m\neq n}^{N}\left|\bm{h}_{n}^{H}\bm{w}_{m}\right|^{2}+\sigma_{n}^{2}}.
\end{equation}
The constant power consumption is averaged by the number of served users in the individual EE due to the fact that all the BSs serve simultaneously all the users. Contrary to conventional communication design approaches, which usually focus on maximizing the spectral efficiency or maximizing the minimum SINR with a maximum transmit power constraint~\cite{JSACZhang2010,TSPBogale2012,TWCHuang2012}, we focus on a fairness-based EE problem to guarantee the EE of each individual node.  In particular, we propose to maximize the minimum user EE, given by
\begin{equation}\label{Fairness_Energy_Efficiency_5}
\max_{\bm{W}}\min_{n}f_{n}\left(\{\bm{w}_{n}\}\right)
s.t.~ tr\left(\bm{B}_{k}\sum_{n=1}^{N}\bm{w}_{n}\bm{w}_{n}^{H}\right) \leq P_{k},\forall k,
\end{equation}
where $\bm{B}_{k}=diag\Big(\underbrace{0,\cdots,0}_{\sum\limits_{m=1}^{k-1}M_{m}},\underbrace{1,\cdots,1}_{M_{k}},
\underbrace{0,\cdots,0}_{\sum\limits_{m=k+1}^{K}M_{m}}\Big)$ and $P_{k}$ denote respectively the transmit power constraint matrix and the individual power constraint of the $k$-th BS, $\bm{W}=\left[\bm{w}_{1},\cdots,\bm{w}_{N}\right]$ denotes the cascaded beamforming matrix. In order to further investigate the relationship between the spectral efficiency and the EE, the conventional minimum user rate maximization problem is also considered, given as
\begin{equation}\label{Fairness_Energy_Efficiency_6}
\max_{\bm{W}}\min_{n}~r_{n}~
s.t.~ tr\left(\bm{B}_{k}\sum_{n=1}^{N}\bm{w}_{n}\bm{w}_{n}^{H}\right) \leq P_{k}, \forall k.
\end{equation}
We treat interference as noise and consider linear receive strategy so that the estimated signal is given by $\tilde{x}_{n}=\mu_{n}y_{n}$. Thus, the MSE of the $n$-th user is given by
\begin{equation}\label{Fairness_Energy_Efficiency_7}
\begin{split}
&\mbox{mse}_{n}=\mathbb{E}\left\{\left(\tilde{x}_{n}-x_{m}\right)\left(\tilde{x}_{n}-x_{m}\right)^{*}\right\}\\
&=\left|\mu_{n}\right|^{2}\sum_{m\neq n}\left|\bm{h}_{n}^{H}\bm{w}_{m}\right|^{2}
+|\mu_{n}|^{2}\sigma_{n}^{2}+\left|1-\mu_{n}\bm{h}_{n}^{H}\bm{w}_{n}\right|^{2}.
\end{split}
\end{equation}
Fixing all the transmit beamformers and minimizing MSE lead to the well-known MMSE receiver:
\begin{equation}\label{Fairness_Energy_Efficiency_8}
\mu_{n}^{opt}=\frac{\bm{w}_{n}^{H}\bm{h}_{n}}
{\sum\limits_{m}\left|\bm{h}_{n}^{H}\bm{w}_{m}\right|^{2}+\sigma_{n}^{2}}.
\end{equation}
and the minimum MSE of the $n$-th user is given by
\begin{equation}\label{Fairness_Energy_Efficiency_9}
\mbox{mse}_{n}^{opt}=1-\frac{\left|\bm{h}_{n}^{H}\bm{w}_{n}\right|^{2}}
{\sum\limits_{m}\left|\bm{h}_{n}^{H}\bm{w}_{m}\right|^{2}+\sigma_{n}^{2}}.
\end{equation}
In what follows, the relationship between the user rate and the user MMSE will be exploited to find the solution to (\ref{Fairness_Energy_Efficiency_5}).

\section*{\sc \uppercase\expandafter{\romannumeral4}. Max-Min Energy Efficiency Algorithm}

It is well known that problem~(\ref{Fairness_Energy_Efficiency_5}) and (\ref{Fairness_Energy_Efficiency_6}) are nonconvex and therefore difficult to solve directly. Furthermore, the fractional form in the objective function (\ref{Fairness_Energy_Efficiency_5}) makes the problem more intractable. To address this issue, we first transform problem~(\ref{Fairness_Energy_Efficiency_5}) into a more tractable form to facilitate the energy efficient algorithm design and then develop an iterative solution. To proceed, we first present the following  equivalent form of problem~(\ref{Fairness_Energy_Efficiency_5}) using the relationship between the user rate and the user MMSE~\cite{TSPShi2011}, given by
\begin{equation}\label{Fairness_Energy_Efficiency_10}
\begin{split}
& \max_{\bm{W}}\min_{n}\max_{\bm{s}, \bm{\mu}} \frac{-s_{n}\mbox{mse}_{n}+\log{s_{n}}+1}{\left\|\bm{w}_{n}\right\|^{2}+\frac{MP_{c}+KP_{0}}{N}}\\
s.t.&~ tr\left(\bm{B}_{k}\sum_{n=1}^{N}\bm{w}_{n}\bm{w}_{n}^{H}\right) \leq P_{k}, \forall k,
\end{split}
\end{equation}
where $\bm{\mu}=\left[\mu_{1},\cdots,\mu_{N}\right]$ and $\bm{s}=\left[s_{1},\cdots,s_{N}\right]$. We have the following lemma which can be proven with a similar method as was used in~\cite{TSPWang2010}.
\begin{lemma}\label{MinMaxLemma}
The minimax equality
\begin{equation}\label{Fairness_Energy_Efficiency_11}
\begin{split}
&\min_{n}\max_{\bm{s}, \bm{\mu}} \frac{-s_{n}\mbox{mse}_{n}+\log{s_{n}}+1}{\left\|\bm{w}_{n}\right\|^{2}+\frac{MP_{c}+KP_{0}}{N}}\\
=&\max_{\bm{s}, \bm{\mu}}\min_{n} \frac{-s_{n}\mbox{mse}_{n}+\log{s_{n}}+1}{\left\|\bm{w}_{n}\right\|^{2}+\frac{MP_{c}+KP_{0}}{N}}
\end{split}
\end{equation}
holds.
\end{lemma}
Based on the above lemma, problem~(\ref{Fairness_Energy_Efficiency_10}) can be rewritten as
\begin{equation}\label{Fairness_Energy_Efficiency_12}
\begin{split}
& \max_{\bm{W}, \bm{s}, \bm{\mu}}\min_{n} \frac{-s_{n}\mbox{mse}_{n}+\log{s_{n}}+1}{\left\|\bm{w}_{n}\right\|^{2}+\frac{MP_{c}+KP_{0}}{N}}\\
s.t.&~ tr\left(\bm{B}_{k}\sum_{n=1}^{N}\bm{w}_{n}\bm{w}_{n}^{H}\right) \leq P_{k}, \forall k.
\end{split}
\end{equation}
It is easily known that problem~(\ref{Fairness_Energy_Efficiency_12}) belongs to a classical generalized fractional programming problem which has been extensively investigated. By applying the fractional theorem,  problem~(\ref{Fairness_Energy_Efficiency_12}) can be written into the following parameterized quadratic subtractive form~\cite{JstorJagan1966,MathCrouzeix1991}
\begin{equation}\label{Fairness_Energy_Efficiency_13}
\begin{split}
& g\left(\eta\right)=\max_{\bm{W}, \bm{s}, \bm{\mu}}\min_{n}~g_{n}\left(\eta\right)\\
s.t.&~ tr\left(\bm{B}_{k}\sum_{n=1}^{N}\bm{w}_{n}\bm{w}_{n}^{H}\right) \leq P_{k}, \forall k,
\end{split}
\end{equation}
where $\eta$ denotes the EE factor which is defined as the minimum value among all individual EEs and $g_{n}\left(\eta\right)$ is given as follows
\begin{equation}\label{Fairness_Energy_Efficiency_14}
\begin{split}
g_{n}\left(\eta\right)=&-s_{n}\mbox{mse}_{n}+\log{s_{n}}+1\\
&-\eta\left(\left\|\bm{w}_{n}\right\|^{2}+\frac{MP_{c}+KP_{0}}{N}\right)\\
=&-s_{n}\left|\mu_{n}\right|^{2}\sum_{m\neq n}\left|\bm{h}_{n}^{H}\bm{w}_{m}\right|^{2}
-\eta\left\|\bm{w}_{n}\right\|^{2}\\
&-s_{n}\left|1-\mu_{n}\bm{h}_{n}^{H}\bm{w}_{n}\right|^{2}+\log{s_{n}}\\
&+1-s_{n}|\mu_{n}|^{2}\sigma_{n}^{2}-\frac{\eta}{N}\left(MP_{c}+KP_{0}\right).
\end{split}
\end{equation}
For a fixed $\bm{W}$, the optimal solutions of $\mu_{n}$ and $s_{n}$ are given with~(\ref{Fairness_Energy_Efficiency_8}) and $s_{n}=\frac{1}{\mbox{mse}_{n}^{opt}}$, respectively. Introducing a slacking variable $\tau$, problem~(\ref{Fairness_Energy_Efficiency_10}) can be reformulated as follows for given $\bm{s}$ and $\bm{\mu}$.
\begin{equation}\label{Fairness_Energy_Efficiency_15}
\begin{split}
& g\left(\eta\right)=\min_{\bm{W}, \tau}~\tau\\
s.t.&~tr\left(\bm{B}_{k}\sum_{n=1}^{N}\bm{w}_{n}\bm{w}_{n}^{H}\right) \leq P_{k}, \forall k.\\
& -g_{n}\left(\eta\right)\leq\tau, \forall n.
\end{split}
\end{equation}
It is easy to see that problem~(\ref{Fairness_Energy_Efficiency_15}) can be solved by using the second order conic programming (SOCP) method~\cite{TSPWiesel2006}.

In the following, a two-layer iterative optimization algorithm is proposed to solve the problem~(\ref{Fairness_Energy_Efficiency_5}). In the outer layer, the EE factor $\eta$ is updated. In the inner layer, the beamformers, receivers and auxiliary variables are updated, respectively. The detailed steps are summarized in Algorithm~\ref{EnergyEfficiency:Solution}.
\begin{algorithm}[H]
\caption{Fairness Energy Efficient Algorithm}\label{EnergyEfficiency:Solution}
\begin{algorithmic}[1]
\STATE Let $t=0$, which denotes the number of iteration, choose arbitrarily $\bm{W}^{\left(t\right)}$ such that it satisfies the power constraints, and compute $\bm{\mu}^{\left(t\right)}$ and $\bm{s}^{\left(t\right)}$.\label{EnergyEfficiency:Initial}
\STATE Let $\tau^{(t)}=0$ and
$\eta^{(t)}=\min\limits_{n} \frac{-s_{n}^{(t)}\mbox{mse}_{n}^{(t)}+\log{s_{n}^{(t)}}+1}
{\left\|\bm{w}_{n}^{(t)}\right\|^{2}+\frac{MP_{c}+KP_{0}}{N}}$.
\label{EnergyEfficiency:UpdatedEta}
\STATE Solve problem~(\ref{Fairness_Energy_Efficiency_15}) with $\bm{\mu}^{(t)}$ and $\bm{s}^{(t)}$, then obtain $\bm{W}^{(*)}$.\label{EnergyEfficiency:UpdatedW}
\STATE Update $\bm{\mu}$ with~(\ref{Fairness_Energy_Efficiency_8}) and $\bm{W}^{(*)}$, then obtain $\bm{\mu}^{(*)}$.\label{EnergyEfficiency:UpdatedU}
\STATE Update $\bm{s}$ with $\bm{W}^{(*)}$ and $\bm{\mu}^{(*)}$, then obtain $\bm{s}^{(*)}$.\label{EnergyEfficiency:UpdatedS}
\STATE Let $\tau^{(*)}=\min\limits_{n}g_{n}\left(\eta^{(t)}\right)$ with $\bm{W}^{(*)}$, $\bm{\mu}^{(*)}$, and $\bm{s}^{(*)}$. If $\left|\tau^{(t)}-\tau^{(*)}\right|\leq\varepsilon$, where $\varepsilon$ is an arbitrarily small positive number, let $\bm{W}^{(t+1)}=\bm{W}^{(*)}$, $\bm{\mu}^{(t+1)}=\bm{\mu}^{(*)}$, $\bm{s}^{(t+1)}=\bm{s}^{(*)}$, $\tau^{(t+1)}=\tau^{(*)}$,  and go step~\ref{Outer:stop}, otherwise let $\bm{W}^{(t)}=\bm{W}^{(*)}$, $\bm{\mu}^{(t)}=\bm{\mu}^{(*)}$, $\bm{s}^{(t)}=\bm{s}^{(*)}$, $\tau^{(t)}=\tau^{(*)}$, and go step~\ref{EnergyEfficiency:UpdatedW}.\label{Inner:stop}
\STATE If $\left|g\left(\eta^{(t)}\right)\right|\leq\epsilon $, where $\epsilon$ is an arbitrarily small positive number, then stop, otherwise let $t=t+1$ and go to step~\ref{EnergyEfficiency:UpdatedEta}.\label{Outer:stop}
\end{algorithmic}
\end{algorithm}

The convergence of Algorithm~\ref{EnergyEfficiency:Solution} can be proven by using a similar method as was used in~\cite{JstorJagan1966,MathCrouzeix1991}. It is worthy noting that this algorithm can be modified to solve the max-min user rate optimization problem~(\ref{Fairness_Energy_Efficiency_6}), starting with rewriting problem~(\ref{Fairness_Energy_Efficiency_6}) into an equivalent form given as
\begin{equation}\label{Fairness_Energy_Efficiency_17}
\begin{split}
& \max_{\bm{W},\bm{\mu},\bm{s}}\min_{n}-s_{n}\mbox{mse}_{n}+\log{s_{n}}+1\\
s.t.&~ tr\left(\bm{B}_{k}\sum_{n=1}^{N}\bm{w}_{n}\bm{w}_{n}^{H}\right) \leq P_{k}, \forall k.\\
\end{split}
\end{equation}
Similar results are observed for this problem. That is, for a fixed $\bm{W}$, the optimal solutions of $\mu_{n}$ and $s_{n}$ are given with~(\ref{Fairness_Energy_Efficiency_8}) and $s_{n}=\frac{1}{\mbox{mse}_{n}^{opt}}$, respectively.  Problem~(\ref{Fairness_Energy_Efficiency_17}) for given $\bm{s}$ and $\bm{\mu}$ can be solved with SOCP. As a consequence, problem~(\ref{Fairness_Energy_Efficiency_17}) can be solved similar to Algorithm~\ref{EnergyEfficiency:Solution}, summarized as Algorithm~\ref{UserRate:Solution}.
\begin{algorithm}[H]
\caption{Fairness User Rate Algorithm}\label{UserRate:Solution}
\begin{algorithmic}[1]
\STATE Let $t=0$, which denotes the number of iteration, choose arbitrarily $\bm{W}^{\left(t\right)}$ such that it satisfies the power constraints, $\bm{\mu}^{(t)}=\bm{0}$, $\bm{s}^{(t)}=\bm{0}$, and let $r^{(t)}=0$.\label{UserRate:Initial}
\STATE Update $\bm{\mu}$ with~(\ref{Fairness_Energy_Efficiency_8}) and $\bm{W}^{(t)}$, then obtain $\bm{\mu}^{(t)}$.\label{UserRate:UpdatedU}
\STATE Update $\bm{s}$ with $\bm{W}^{(t)}$ and $\bm{\mu}^{(t)}$, then obtain $\bm{s}^{(t)}$.\label{UserRate:UpdatedS}
\STATE Solve problem (\ref{Fairness_Energy_Efficiency_17}) with SOCP method for given $\bm{\mu}^{(t)}$, and $\bm{s}^{(t)}$, then obtain $\bm{W}^{(t)}$.\label{UserRate:UpdatedW}
\STATE Let $r^{(t)}=\min\limits_{n}-s_{n}\mbox{mse}_{n}+\log{s_{n}}+1$ with $\bm{W}^{(t)}$, $\bm{\mu}^{(t)}$, and $\bm{s}^{(t)}$. If $\left|r^{(t+1)}-r^{(t)}\right|\leq\varepsilon$, where $\varepsilon$ is an arbitrarily small positive number, let $\bm{W}^{(t+1)}=\bm{W}^{(t)}$, $r^{(t+1)}=\tau^{(t)}$, and stop, otherwise let $\bm{W}^{(t+1)}=\bm{W}^{(t)}$, $t=t+1$, and go step~\ref{UserRate:UpdatedU}.
\end{algorithmic}
\end{algorithm}
\begin{remark}
\rm Algorithm~\ref{EnergyEfficiency:Solution} and Algorithm~\ref{UserRate:Solution} can be easily extended to the multicell multiuser coordinated beamforming case where the BSs only cooperate in beamforming design and need no data sharing.
\end{remark}

\section*{\sc \uppercase\expandafter{\romannumeral5}. Numerical Results}

In this section, the performance of the proposed multicell beamforming schemes is investigated via numerical simulations. We consider a cluster of $K$ cooperating BSs. The inter-BS distance is $1$km and each user has at least $400$m distance from its serving BS. The small scale fading channel coefficient $\bm{h}_{i,j}$ is assumed to be Gaussian distributed with zero mean and identity covariance matrix, $\eta_{i,j}$ denotes the large scale fading factor which in decibels was given as $10\log_{10}(\eta_{i,j})=-38\log_{10}(d_{i,j})-34.5+\varsigma_{i,j}$, where $\varsigma_{i,j}$ represents the shadow fading in decibels and follows the distribution $\cal N$($0$,$8$dB). The transmit power budget is set to $P$ for each BS and the SNR in the figures is defined as the transmit power in decibel, i.e., $\rm{SNR}=10\log_{10}\rm{P}$. The noise figure at each user terminal is $9$dB. Assuming that each BS has the same power constraint over $10$MHz bandwidth and $\epsilon=10^{-4}$.

\begin{figure}[h]
\centering
\captionstyle{flushleft}
\onelinecaptionstrue
\includegraphics[width=1\columnwidth,keepaspectratio]{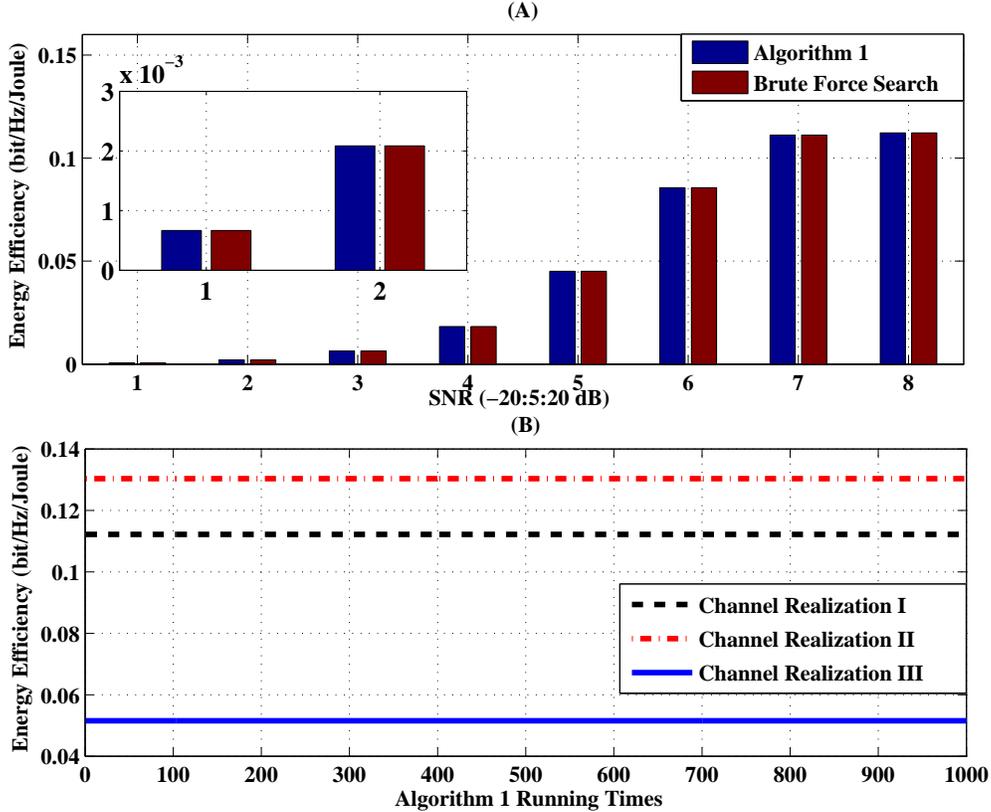}\\
\caption{EE of Algorithm~\ref{EnergyEfficiency:Solution} under different transmit power constraint for arbitrary channel realization, $P_{c}=30$dBm, $P_{0}=40$dBm, $K=3$, $\widetilde{M}=4$, and $\widetilde{N}=1$.}
\label{FairnessEnergyEfficiencyOptimal}
\end{figure}

Sub-figure (A) in Fig.~\ref{FairnessEnergyEfficiencyOptimal} compares the EE performance of our solution with the optimal performance achieved by brute force search. Numerical results show that the proposed algorithm achieves a performance very close to the global optimum. To further examine the impact of beamforming initialization on the performance of the proposed algorithm, Sub-figure (B) in Fig.~\ref{FairnessEnergyEfficiencyOptimal} shows the fairness of the EE performance of Algorithm~\ref{EnergyEfficiency:Solution} varying with algorithm running times each with independent random initialization, over a few random channel realizations with $\rm{SNR}=15$dB. Numerical results demonstrate that our proposed algorithm always achieves the same performance with arbitrary random beamforming initialization.

\begin{figure}[h]
\centering
\captionstyle{flushleft}
\onelinecaptionstrue
\includegraphics[width=1\columnwidth,keepaspectratio]{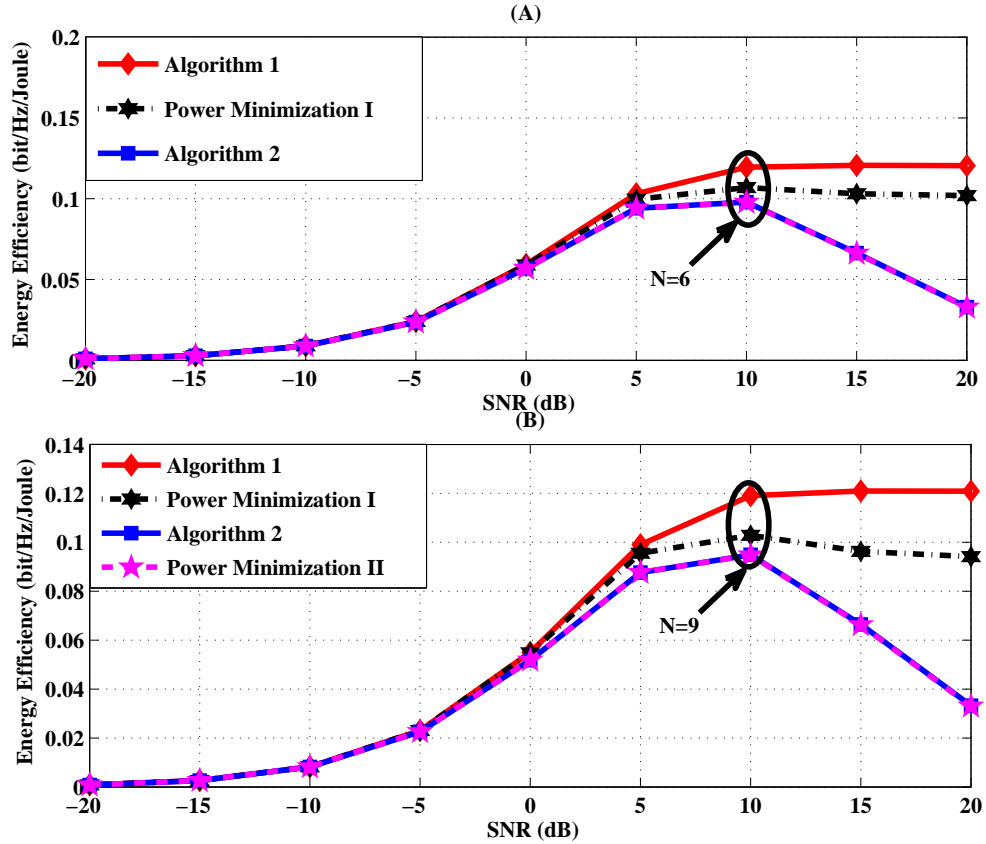}\\
\caption{Minimum EE comparison for different algorithms, $P_{c}=30$dBm, $P_{0}=40$dBm, $K=3$, and $\widetilde{M}=4$.}
\label{FairnessEnergyEfficiencyComparison}
\end{figure}

Fig.~\ref{FairnessEnergyEfficiencyComparison} illustrates the average minimum EE of several algorithms over $10000$ random channel realizations. Power Minimization \uppercase\expandafter{\romannumeral1} (\uppercase\expandafter{\romannumeral2}) denotes the minimum user EE achieved by the total power minimization scheme which can be solved by in~\cite{TWCDahrouj2010} where the target user rate is set as the corresponding user rate achieved by Algorithm~\ref{EnergyEfficiency:Solution} (Algorithm~\ref{UserRate:Solution}). To guarantee fairness, the power minimization scheme was solved by applying the SOCP method~\cite{TSPWiesel2006}. Numerical results show that all the algorithms achieve the same EE performance in the lower SNR region such as $-20\sim -5$dB. While in the middle-high SNR region such as $-5\sim 20$dB, Algorithm~\ref{EnergyEfficiency:Solution} obtains a better performance than other three algorithms in terms of maximizing the minimum EE. Since the power minimization scheme only minimizes total power consumption for given rate targets, it is shown that Power Minimization \uppercase\expandafter{\romannumeral1} exhibits lower fairness EE than Algorithm~\ref{EnergyEfficiency:Solution} in the middle-high SNR region, while Power Minimization \uppercase\expandafter{\romannumeral2} and Algorithm~\ref{UserRate:Solution} always achieve the same EE performance in the whole SNR region.


\section*{\sc \uppercase\expandafter{\romannumeral6}. Conclusion}

In this letter, an EE optimization problem which takes the fairness among users into account was considered. In order to solve the problem, the relationship between the user rate and the MMSE was first used to rewrite the user rate into a tractable form. Then, the fractional optimization problem was transformed into a parameterized polynomial subtractive form by applying the fractional programming theorem. Based on that, an iterative algorithm with proved convergence was proposed to solve the max-min EE optimization problem.

\begin{small}

\end{small}

\end{document}